\documentclass[bibyear%
]{an}
\usepackage{graphicx}
\usepackage{times}

\def\HI{H{\,\small I}}
\def\Htwo{H$_{\,2}$}

\newcommand{\Htwoi}{H$_{\,2}$ 1-0 S(1)}
\newcommand{\Htwot}{H$_{\,2}$ 1-0 S(3)}

\newcommand{\kms}{$\,$km$\,$s$^{-1}$}

\newcommand{\msun}{{${\rm M}_\odot$}}

\def\HI{H{\,\small I}}

\def\emph#1{{\sl #1}}
\newcommand{\ltsima} {$\; \buildrel < \over \sim \;$}
\newcommand{\gtsima} {$\; \buildrel > \over \sim \;$}
\newcommand{\lta} {\lower.5ex\hbox{\ltsima}}
\newcommand{\gta} {\lower.5ex\hbox{\gtsima}}

\newcommand{\pks}{\mbox PKS B1718-649}

\begin{document}

\Pagespan{1}{}
\Yearpublication{2011}%
\Yearsubmission{2011}%
\Month{1}%
\Volume{999}%
\Issue{92}%

\title{\pks: an \HI\ and \Htwo\ perspective on the birth of a compact
radio source.}

\author{F. M. Maccagni\inst{1}$^,$\inst{2}\fnmsep\thanks{Corresponding author:
  \email{maccagni@astro.rug.nl}}
   \and F. Santoro\inst{1}$^,$\inst{2}
   \and R. Morganti\inst{1}$^,$\inst{2}
   \and T. A. Oosterloo\inst{1}$^,$\inst{2} 
   \and J. B. R. Oonk\inst{2}$^,$\inst{3} 
   \and B. H. C. Emonts\inst{4}}

   \institute{Kapteyn Astronomical Institute, Rijksuniversiteit Groningen, Landleven 12, 9747 AD Gronignen, The Netherlands
   \and Netherlands Institute for Radio Astronomy, Postbus 2, 7990 AA, Dwingeloo, The Netherlands
   \and Leiden Observatory, Leiden University, PO Box 9513, NL-2300 RA Leiden, the Netherland
   \and Centro de Astrobiologi\'ia (INTA-CSIC), Ctra de Torrej\'on a Ajalvir, km 4, 28850 Torrej\'on de Ardoz, Madrid, Spain\\
                   }

\received{XXXX}
\accepted{XXXX}
\publonline{XXXX}

\keywords{PKS B1718-649, compact radio sources, active nuclei, neutral hydrogen, molecular hydrogen, ISM.}

\abstract{We present neutral hydrogen (\HI) and warm molecular hydrogen (\Htwo) observations of the young ($10^2$ years) radio galaxy \pks. We study the morphology and the kinematics of both gas components, focusing, in particular, on their properties in relation to the triggering of the radio activity. The regular kinematics of the large scale \HI\ disk, seen in emission, suggests that an interaction event occurred too long ago to be responsible for the recent triggering of the radio activity. In absorption, we detect two absorption lines along the narrow line of sight of the compact ($r<2$ pc) radio
source. The lines trace two clouds with opposite radial motions. These may represent a population of clouds in the very inner regions of the galaxy, which may be involved in triggering the radio activity. The warm molecular hydrogen (\Htwoi\ ro-vibrational line) in the innermost kilo-parsec of the galaxy appears to be distributed in a circum-nuclear disk following the regular kinematics of the \HI\ and of the stellar component. An exception to this behaviour arises only in the very centre, where a highly dispersed component is detected. These particular \HI\ and \Htwo\ features suggest that a
strong interplay between the radio source and the surrounding ISM is on-going. The physical properties of the cold gas in the proximity of the radio source may regulate the accretion recently triggered in this AGN.}

\maketitle

\section{Introduction}
Compact radio Active Galactic Nuclei (AGN) are often embedded in a dense multiphase Interstellar Medium (O'Dea et al., 1991; Orienti et al., 2007; Pasetto et. al., these proceedings; O'Dea et al., these proceedings). Many compact sources seem to be at their first stages of radio activity (Murgia et al., 1999; Parma et al., 1999), hence they appear young and embedded in a dense environment. Yet, the evolution of radio galaxies is varied. In some sources the radio jet escapes from the host galaxy, while in others, it remains confined in the dense surrounding medium. In some sources, the radio activity ceases at a very early stage (Baum et al., 1990; Fanti, 2009) and in others it is re-activated afterwards (Stanghellini et al., 2005; Brienza et al., these proceedings). Hence, it is plausible that different physical conditions of the dense ISM, in the centre of galaxies, determine the triggering of a radio source and regulate its activity. Among the gaseous components of compact sources, the atomic and molecular is the most massive. It remains unclear if and under which physical conditions the cold gas contributes to the evolution of the radio activity. Nevertheless, the distribution and the kinematics of the cold gas in the host galaxies of the radio sources allow us to trace the interaction history of the galaxy as well as the presence of `secular' processes (Emonts et al.,2006; Struve et al., 2010). These phenomena are often invoked to bring the gas into the inner regions of galaxies, in order to trigger and fuel the radio activity.

Recent surveys (Vermeulen et al., 2003; Orienti et al., 2006; Emonts et al., 2010; Chandola et al., 2013; Ger\'eb et al., 2014; Ger\'eb et al., 2015) show that compact young sources are rich in neutral hydrogen (\HI). The \HI\ detected in absorption against the radio core of these sources has unsettled kinematics with respect to the host galaxy. Some young sources also appear to have abundant molecular hydrogen (\Htwo) (Garc\'ia-Burillo et al., 2007; Willett et al., 2010; Dasyra $\&$ Combes, 2011; Hicks et al., 2013). The kinematics of the molecular hydrogen seems to show regular rotation, while high velocity dispersion rises only in the very inner regions (Guillard et al., 2012, M\"uller-S\'anchez et al., 2013). In such an environment, it is likely that the radio source expands perturbing the neutral and molecular gas. On the other hand, it is also possible that the turbulence of the gas creates the conditions for the triggering of the radio source. Moreover, the way gas is accreted onto the black hole, the `fueling' mechanism, may determine the characteristics of the radio source associated with the AGN (Best $\&$ Heckman, 2012). Hence, the atomic and molecular gas (e.g. \HI\ and \Htwo), in the innermost region of compact sources, may be crucial in triggering a young radio source and driving its evolution.

\pks\ is one of the closest (d $= 62.4$ Mpc), most compact (r$< 2$ pc) and youngest radio sources ($10^2$ years; Tingay et al., 1997; Giroletti $\&$ Polatidis, 2006). Thus, it is an ideal candidate to study with high resolution observations, the kinematics of the hydrogen (atomic and molecular) in proximity of the radio activity. \pks\ can be classified as a LINER (Filippenko, 1985); it has low-efficiency accretion ($\lambda_{\rm Edd}\sim0.003$) and it is low-power in the radio band ($P_{1.4 \rm{GHz}}=1.8\times10^{24}$ W Hz$^{-1}$). The radio source is embedded in a massive \HI\ disk. The analysis of the kinematics of the neutral hydrogen in relation to the triggering of the young radio source has been presented in Maccagni et al. (2014). In this paper, we summarise the main results from the \HI\ observations, and we present new SINFONI-VLT observations of the warm molecular hydrogen in the innermost $2.5$ kpc of the galaxy. Its kinematics provides new insights on the assembly of the molecular gas around the young radio source. Furthermore, we discuss the implications that the physical conditions of the cold gas can have for fuelling a low-power radio source, such as \pks.

\begin{figure}
\begin{center}
\includegraphics[trim = 0 0 0 0, clip,width=0.5\textwidth]{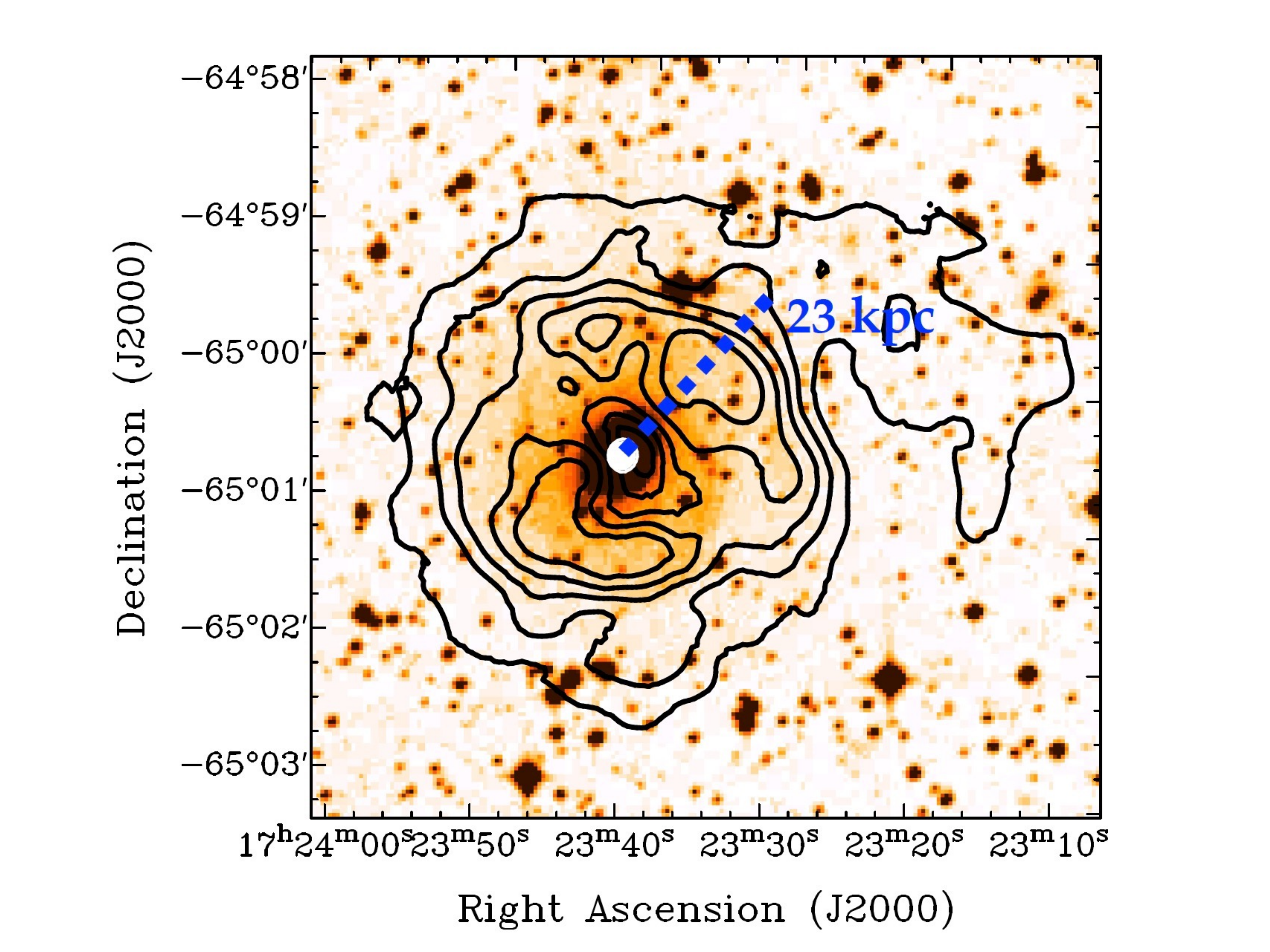}
\caption{I-band optical image of \pks, overlaid with the column density contours (in {\em black}) of the neutral gas. The \HI\ disk has the shape of an incomplete ring with asymmetries in the N-W and in the S of the disk. The contour levels range between $7\times10^{19}$ cm$^{-2}$ and$8\times10^{20}$ cm$^{-2}$, with steps of $1.5\times10^{20}$ cm $^{-2}$. The unresolved continuum radio source is marked in {\em white}. The maximum radius of regular rotation is marked in {\em dashed blue}.}

\label{fig:intensity}
\end{center}
\end{figure}
\section{The \HI\ of \pks}
As seen in many radio sources, \pks\ is hosted by an early-type galaxy, \mbox{NGC 6328}. Nevertheless, it has a massive ($M_{\rm H{\,\small I}} = 1.1\times10^{10}$\msun) face-on \HI\ disk, extending out to $\sim 30$ kpc from the nucleus (as shown in~\ref{fig:intensity}, Veron-C\'etty et al., 1995). Deeper and higher resolution observations were recently taken with the Australian Telescope Compact Array (ATCA) in order to study the kinematics of the \HI\ disk and, consequently, to trace the interaction history of the host galaxy (see Maccagni et al. (2014) for a detailed analysis). These observations are well matched by a warped disk regularly rotating up to the distance of $23$ kpc, where the neutral hydrogen starts deviating from circular motions. This radius (in {\em dashed blue} in Fig.~\ref{fig:intensity}) can be connected to the date of the last interaction, assuming that the \HI\ takes at least two orbits to settle into regular rotation. Using the rotational velocity of the disk ($v_{rot} = 220$\kms), we date the last interaction of \pks\ to $1\times10^9$ years. Given the age of the radio source ($10^2$ years), the connection between an interaction event and the triggering of the radio source is unlikely. Also, down to $3$ kpc (the minimum inner radius where the \HI\ is detected in emission), we do not detect gas significantly deviating from regular rotation, to be moving towards the radio source. 

\begin{figure}
\begin{center}
\includegraphics[trim = 0 0 0 0, clip,width=0.5\textwidth]{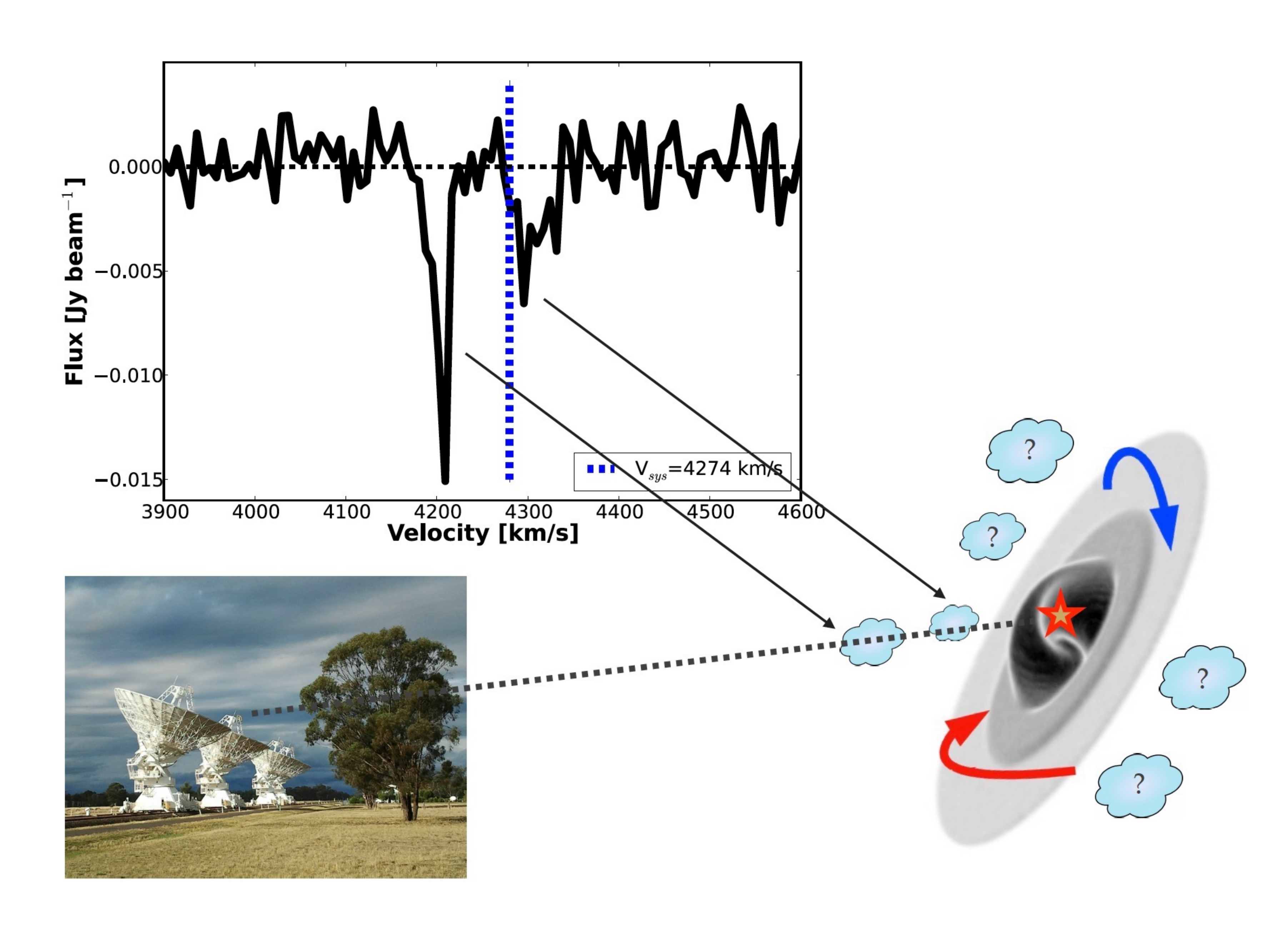}
\caption{{\bf (top)}: \HI\ profile of \pks\ obtained from the ATCA data. The two absorption systems are clearly visible. The narrow line is located at velocity $4200$\kms, blue-shifted with respect to the systemic velocity ($4274$\kms, {\em dashed blue} line). The broad component peaks at $4300$\kms. {\bf (bottom)}: the two lines detected against a very narrow line of sight (r $< 2$ pc). They trace two clouds not regularly rotating within the disk detected in emission. These clouds may belong to a larger population which could be contributing to fuel the nuclear activity}
\label{fig:plot}
\end{center}
\end{figure}

In \pks, the \HI\ is also detected in absorption against the very narrow (r$<2$ pc) radio core. Fig.~\ref{fig:plot}~(top) shows two kinematically-separated absorption lines. The narrow line ($FWZI=43$\kms) is blue-shifted ($\Delta v = v_{\rm peak} - v_{\rm sys}= -75$ \kms; where  $v_{\rm sys} = 4274$ \kms), while the broader ($FWZI = 65$ \kms) one is red-shifted ($v=+26 $\kms) , with respect to the systemic velocity. Given the small size of the background source and the different kinematics of the lines, the \HI\ detected in absorption does not belong to the warped regularly rotating disk, seen in emission. The most likely possibility is that both absorption components come from two small clouds, not regularly rotating (\ref{fig:plot} (bottom)). These clouds must be small ($10^4$~\msun$\leq M_{\rm H{\,\small I}}\leq 10^7$~\msun), otherwise they would have been detected in emission. Although these particular clouds may not be directly interacting with the active nucleus, they may belong to a larger population present in the nuclear region of the galaxy (r $< 3$ kpc), which may contribute to fuel the AGN by falling into the nucleus. Nevertheless, the location of the \HI\ clouds cannot be determined with accuracy. Such a  population close to the radio source is indirectly suggested by the radio continuum variability of \pks\ in the $1-3$ GHz band. This has been explained by variations in the opacity of the ISM due to free-free absorption in a clumpy dense medium, which can be potentially related with an ensemble of clouds in the circum-nuclear region of the radio source (Tingay et al., 2015).

The information collected so far suggests that, in the very inner regions of \pks, the atomic and molecular hydrogen may constitute the reservoir of material actively fuelling the AGN. A better insight into the structure of the circum-nuclear regions, and on the role of the \HI\ clouds, in relation to the fuelling of the radio source, can be given only by high resolution observations of the innermost hundreds of parsecs. This can be done using the molecular hydrogen as tracer.

In AGN similar to \pks\ (e.g \mbox{4C +31.06} and \mbox{NGC 3169}), the nuclear regions often appear be rich in molecular gas (Garc\'ia-Burillo et al., 2007; Struve et al., 2012; M\"uller-S\'anchez et al., 2013 ). The molecular gas often is assembled in a disk-like clumpy structure around the nuclear activity, and could potentially constitute the fuel reservoir of the AGN. By means of integral-field spectroscopy, it has been possible to trace with high detail the distribution of the \Htwo\ around the AGN, study its kinematics and relate these to the fuelling of the nuclear activity (Hicks et al., 2009, 2013; M\"uller-S\'anchez et al., 2009, 2013). In \pks\, molecular hydrogen has been detected in its lowest rotational state (\Htwo\ $0-0$ S(0,...,6)) using Spitzer (Willett et al., 2010). These observations do not have high enough spatial resolution to determine the distribution of the \Htwo\ in the nuclear regions needed to constrain the role of the hydrogen gas in fuelling of the AGN. On the other hand, SINFONI-VLT K-band observations allow us to trace the distribution and kinematics of the \Htwo\ $1-0$ S(0,...,3) ro-vibrational states. The preliminary analysis of the observations is presented in the next Section.

\begin{figure*}
\begin{center}
\includegraphics[trim = 0 0 0 0, clip,width=\textwidth]{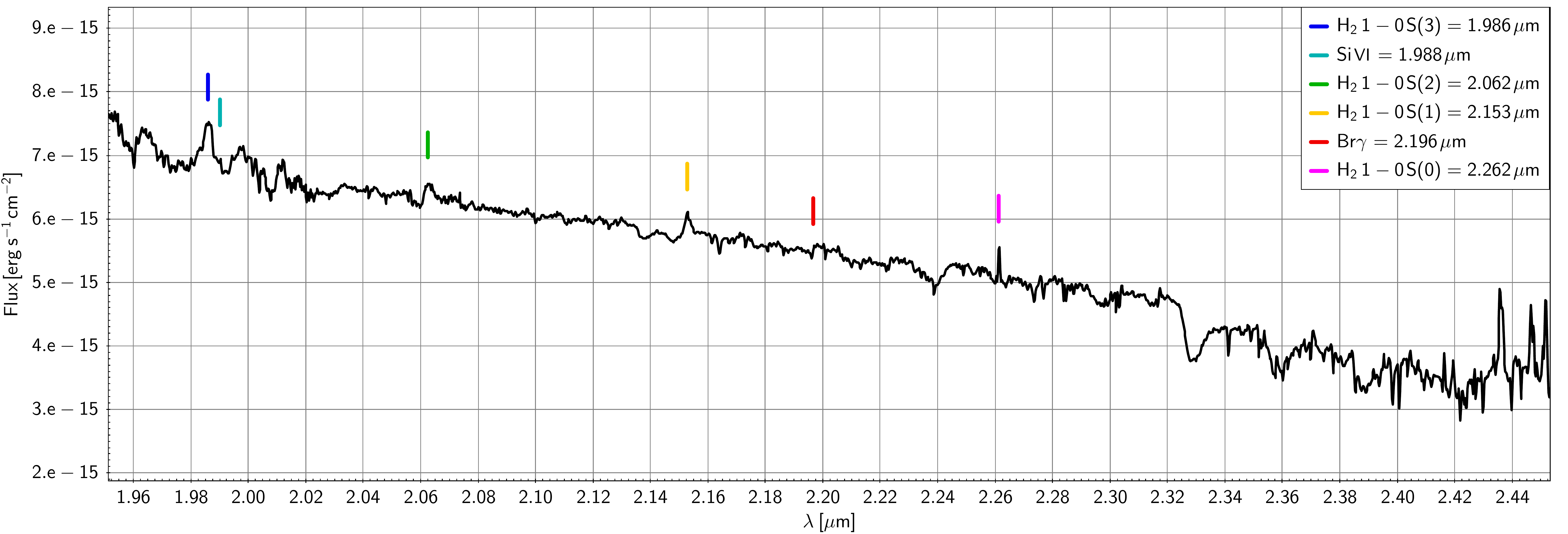}
\caption{Sky-corrected $1.95-2.45\mu$m near-IR spectrum of \pks, extracted in the central $300$ pc. The solid color bars show the \Htwo\ ro-vibrational lines we detect, while the dashed mark the ones we could have detected in the same wavelength range.}
\label{fig:spectrum}
\end{center}
\end{figure*}

\section{The \Htwo\ of \pks}
The SINFONI-VLT instrument has a field of view $8''\times8''$ ($2.5\times2.5$ kpc, at the redshift of \pks). The observations of \pks\ are seeing-limited, reaching a spatial resolution of $0.5''$ ($r\sim200$ pc).  In the K-band we detect $4$ emission lines of the lowest ro-vibrational states of the molecular hydrogen (\Htwo\ $1-0$ S(0,...,3)). The spectrum, extracted in the central $300$ pc of \pks\ is shown in Fig.~\ref{fig:spectrum}. The \Htwoi\ and \Htwot\ are the two brightest lines, clearly detected at $2.15\,\mu$m and $1.98\,\mu$m, respectively

\begin{figure*}
\begin{center}
\includegraphics[trim = 0 0 0 0, clip,width=.49\textwidth]{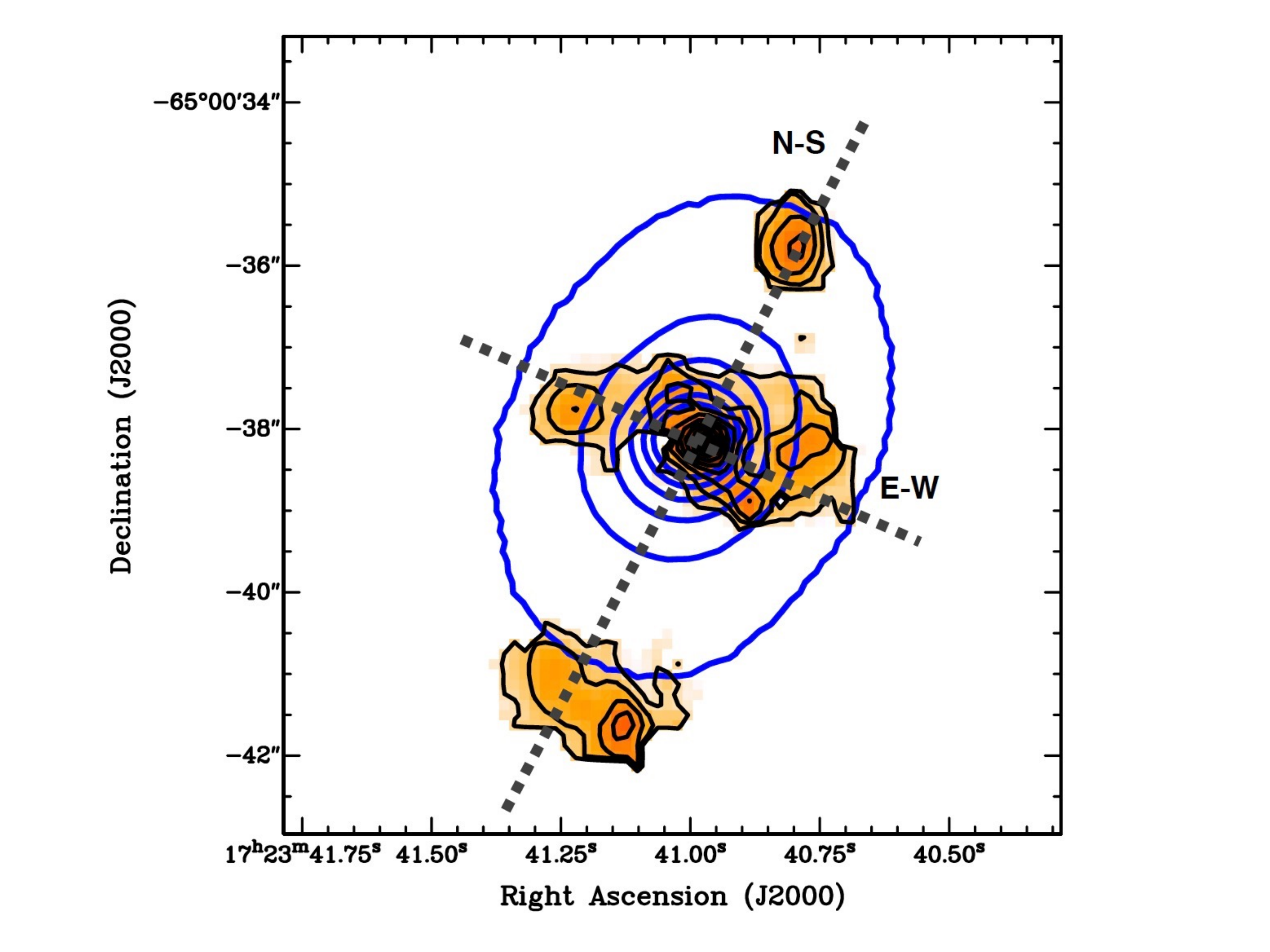}
\includegraphics[trim = 0 0 0 0, clip,width=.49\textwidth]{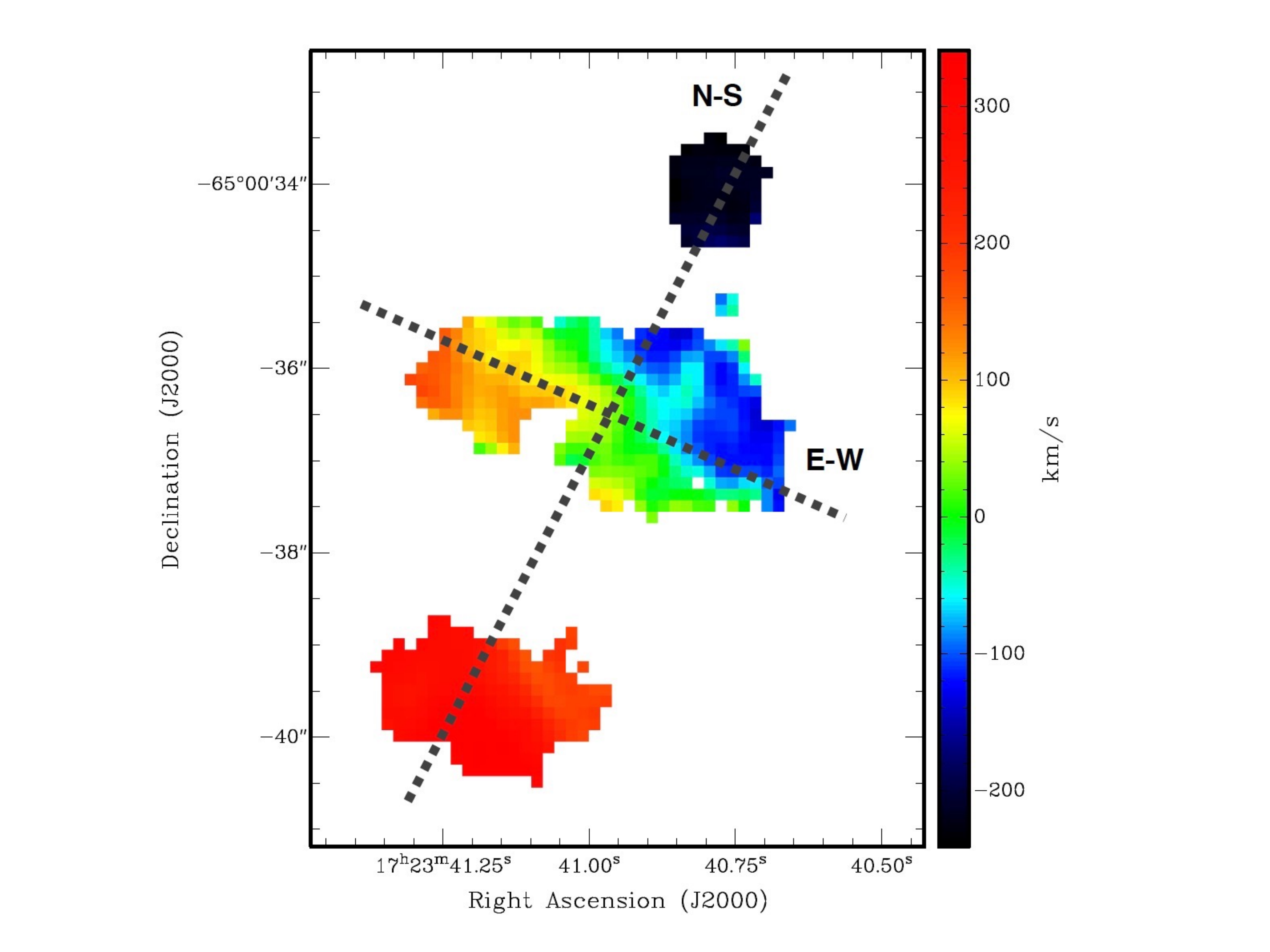}
\caption{ {\bf (a)}: Intensity map of the \Htwoi\ line in the inner $5$ kpc of \pks, overlaid with the isophotes of the stellar emission, (in {\em blue}). The major axis of the outer \Htwo\ (N-S) is coincident to that of the stars and of the large scale \HI\ disk. {\bf (b)}: Velocity field of the warm \Htwo, the colour map is centred at the \HI\ systemic velocity ($v_{sys} = 4274$ \kms). The outer regions follow the rotation of the \HI\ disk, while the inner central region has a different axis of rotation E-W.}
\label{fig:htwo}
\end{center}
\end{figure*}

The complete analysis of the SINFONI observations will be presented in Maccagni et al., (in prep.). In the following paragraphs, we focus on the analysis of the distribution and the kinematics of the \Htwoi\ line. Fig.~\ref{fig:htwo}(a,b) shows that in the outer regions (r $> 2$ kpc), the warm \Htwo\ (T $\sim10^3$ K) follows the regular rotation of the galaxy, matching the velocities observed in the \HI. The major axis is oriented as the stellar disk (blue contours in the image) and as the major axis of the \HI\ disk (approximately along the N-S direction). In the inner regions (r$ < 1$ kpc), the warm \Htwo\ abruptly changes the axis of rotation, forming a circum-nuclear disk oriented E-W. The velocity field shows that overall the \Htwo\ kinematics are supported by rotation, with velocity peaking at $\sim200$\kms. In the very inner regions of the disk (r $< 200$ pc), the velocity dispersion increases abruptly to $250$\kms. Given the close proximity of the radio source, this may be due to a gas component non-rotating withing, but falling towards the AGN, thus contributing to its fuelling. This component of \Htwo\ (with high velocity dispersion) may represent the molecular counterpart of the cloud population detected in \HI\ absorption. If this is the case, the \HI\ would be embedded in the inner circum-nuclear \Htwo\ disk and thus, the disk surrounding the radio source would be clumpy and multi-phase, such as the one proposed to explain the variability of the $1-3$ GHz spectrum (Tingay et al., 2015). In this scenario, i.e. \HI\ clouds embedded in the \Htwo\ circum-nuclear medium, variability of the the optical depth of the \HI\ absorption lines should also be found over timescales of months to years. Measuring such variability could set further constraints on the structure and density of the atomic and molecular medium around the radio source.

The joint analysis of the kinematics of the \HI\ and the ro-vibrational line of the \Htwo\ strongly suggests that in \pks\ the neutral and molecular hydrogen, in the inner $200$ pc, have kinematics deviating from regular rotation. Given the small size of the radio source (r $< 2$ pc) it is unlikely that during its expansion the radio source has perturbed the gas at distances of hundreds of parsecs. On the other hand, it is plausible that the turbulent gas is feeding the radio activity, as has been proposed to explain the similar properties of the \Htwo\ in Seyfert galaxies (Hicks et al., 2013). Moreover, given the low-efficiency accretion nature of \pks\ ($\lambda_{\rm Edd}\sim0.003$, Maccagni et al., 2014), advection-dominated mechanisms in an optically thick environment should be the typical accretion for this AGN (Best $\&$ Heckman, 2012). Hence, it is plausible to interpret the peculiar kinematics of the cold gas as the effects of clouds falling into the AGN and actively contributing to its fuelling.

As previously metioned, \pks\ has been classified as a LINER galaxy (Filippenko, 1985). There are a few other LINER galaxies (e.g. \mbox{NGC 3169}, \mbox{NGC 1052}, \mbox{NGC 2911}), where the warm \Htwo\ has been studied in detail close to the AGN (M\"uller-S\'anchez et al., 2013). In these objects, like in \pks, the \Htwo\ is distributed into a circum-nuclear disk with high velocity dispersion ($\sigma\sim100$\kms) within $100$ pc from the nucleus, suggesting that the molecular gas may form a clumpy circum-nuclear disk around the AGN and actively fuel it.  In these galaxies, the Si$[VI]$ line is non detected, as in \pks. This may hint that some compact sources are rich in cold gas (molecular and neutral) but the accretion mechanism is not powerful enough to produce hard photons able to ionise the coronal region. This is consistent with the low-efficiency accretion of \pks, mentioned above. Other compact radio sources show properties in the neutral and molecular gas similar to \pks. In particular, \mbox{4C +31.06} has a molecular gas disk rotating with the host galaxy. Nevertheless, in the nuclear regions both the \Htwo\ and the \HI\ show strong non-circular motions (Garc\'ia-Burillo et al., 2007, Struve et al., 2010).

\section{Conclusions}
In this paper we presented the neutral and molecular hydrogen observations of the compact young radio source \pks. The kinematics of the large scale \HI\ disk allows us to exclude an interaction event or a `secular' process as responsible for the triggering and fuelling of this radio source. The origin of such mechanisms must instead be found in the very inner regions of this galaxy. The \HI\ absorption lines, given their opposite velocities with respect to the systemic velocity of the gas, suggest that a population of gas clouds may surround the radio source, eventually contributing to its fuelling. This scenario is further supported by the kinematics of the \Htwo. In the innermost kilo-parsec, the \Htwo\ is distributed in a regularly rotating circum-nuclear disk. An exception to this arises only within $200$ pc from the radio source, where the velocity dispersion steeply increases. A further hint on the presence around \pks\ of a clumpy circum-nuclear turbulent medium, which can potentially fuel the radio activity, is provided by the variability of the $1-3$ GHz spectrum (Tingay et al. 2015).

Atacama Large Millimeter Array (ALMA) high resolution observations of the cold phase of the molecular gas (e.g. traced by CO) can provide further insights on the distribution and kinematics of the gas. ALMA would be the ideal instrument to observe the molecular gas in \pks\ on scales of tens of parsecs. This would certainly allow us to take a leap forward in the understanding of the kinematics of the molecular gas in the proximity of young radio sources, and its contribution to their fuelling.

The observation of the innermost $100$ pc of radio sources, like in this case \pks, often shows that the gas in both the neutral and the molecular phase has unsettled structures which may be fuelling the radio activity. The presence of such structures, in the nuclear regions is in agreement with the general properties of the cold gas in compact sources, which have been described by shallow observations of statistical samples of such sources (e.g. Emonts et al., 2010; Chandola et al., 2013; Ger\'eb et al., 2014; Ger\'eb et al., 2015). High resolution observations, which determine the distribution of the \HI\ in the nuclear regions, can be performed using the Very Large Baseline Interferometer (VLBI). The large number of young radio sources known by now to have \HI\ with unsettled components (Curran et al., 2013; Ger\'eb et al., 2014; Ger\'eb et al., 2015) allows us to develop these studies over a statistical sample of sources. Some of these studies have already begun, as for example the VLBI observations with of \mbox{NGC 315}, \mbox{4C +12.50}, \mbox{3C 293}, \mbox{4C +52.37} (Beswick et al., 2004; Morganti et al., 2004, 2009, Nyland et al. in prep.).

\acknowledgements
The research leading to these results has received funding from the European Research Council under the European Union's Seventh Framework Programme (FP/2007-2013) / ERC Advanced Grant RADIOLIFE-320745. BE acknowledges funding by the European Union 7th Framework Programme (FP7-PEOPLE-2013-IEF) grant 624351. The authors would like to thank the organisers and RadioNet for the support to this inspiring workshop. 

%
 \bibliographystyle{an/an}
\bibliography{biblio}

\begin{thebibliography}{}

\bibitem{baum}
{Baum}, S.\,A., {O'Dea}, C.\,P., {Murphy}, D.\,W., {de Bruyn}, A.\,G.: 1990,
  A\&A 232, 19--26

\bibitem{best}
{Best}, P.\,N., {Heckman}, T.\,M.: 2012,  MNRAS 421, 1569--1582

\bibitem{beswick}
{Beswick}, R.\,J., {Peck}, A.\,B., {Taylor}, G.\,B., {Giovannini}, G.: 2004,
  \mnras 352, 49--60

\bibitem{chandola2013}
{Chandola}, Y., {Gupta}, N., {Saikia}, D.\,J.: 2013,  \mnras 429, 2380--2391

\bibitem{curran}
{Curran}, S.\,J., {Allison}, J.\,R., {Glowacki}, M., {Whiting}, M.\,T.,
  {Sadler}, E.\,M.: 2013,  \mnras 431, 3408--3413

\bibitem{dasyra}
{Dasyra}, K.\,M., {Combes}, F.: 2011,  A\&A 533, L10

\bibitem{emonts}
{Emonts}, B.\,H.\,C., {Morganti}, R., {Struve}, C., {Oosterloo}, T.\,A., {van
  Moorsel}, e.\,a.: 2010,  MNRAS 406, 987--1006

\bibitem{emonts1}
{Emonts}, B.\,H.\,C., {Morganti}, R., {Tadhunter}, C.\,N., {Holt}, J.,
  {Oosterloo}, e.\,a.: 2006,  A\&A 454, 125--135

\bibitem{fanti}
{Fanti}, C.: 2009,  Astronomische Nachrichten 330, 120--127

\bibitem{filippenko}
{Filippenko}, A.\,V.: 1985,  ApJ 289, 475--489

\bibitem{garcia2007}
{Garc{\'{\i}}a-Burillo}, S., {Combes}, F., {Neri}, R., {Fuente}, A., {Usero},
  A., {Leon}, S., {Lim}, J.: 2007,  A\&A 468, L71--L75

\bibitem{gereb2}
{Ger{\'e}b}, K., {Maccagni}, F.\,M., {Morganti}, R., {Oosterloo}, T.\,A.: 2015,
   A\&A 575, A44

\bibitem{gereb1}
{Ger{\'e}b}, K., {Morganti}, R., {Oosterloo}, T.\,A.: 2014,  A\&A 569, A35

\bibitem{giroletti}
{Giroletti}, M., {Polatidis}, A.: 2006,  Astron. Nachr. 88, 789--794

\bibitem{guillard}
{Guillard}, P., {Ogle}, P.\,M., {Emonts}, B.\,H.\,C., {Appleton},
  P.\,N.\,e.\,a.: 2012,  \apj 747, 95

\bibitem{hicks2013}
{Hicks}, E.\,K.\,S., {Davies}, R.\,I., {Maciejewski}, W., {Emsellem}, E.,
  {Malkan}, e.\,a.: 2013,  \apj 768, 107

\bibitem{hicks}
{Hicks}, E.\,K.\,S., {Davies}, R.\,I., {Malkan}, M.\,A., {Genzel}, R.,
  {Tacconi}, e.\,a.: 2009,  ApJ 696, 448--470

\bibitem{maccagni}
{Maccagni}, F.\,M., {Morganti}, R., {Oosterloo}, T.\,A., {Mahony}, E.\,K.:
  2014,  A\&As 571, A67

\bibitem{morganti2004}
{Morganti}, R., {Oosterloo}, T.\,A., {Tadhunter}, C.\,N., {Vermeulen}, R.,
  {Pihlstr{\"o}m}, Y.\,M., {van Moorsel}, G., {Wills}, K.\,A.: 2004,  A\&A 424,
  119--124

\bibitem{morganti2009}
{Morganti}, R., {Peck}, A.\,B., {Oosterloo}, T.\,A., {van Moorsel}, G.,
  {Capetti}, e.\,a.: 2009,  A\&A 505, 559--567

\bibitem{muller2009}
{M{\"u}ller-S{\'a}nchez}, F., {Davies}, R.\,I., {Genzel}, R., {Tacconi},
  L.\,J., {Eisenhauer}, F., {Hicks}, e.\,a.: 2009,  \apj 691, 749--759

\bibitem{muller}
{M{\"u}ller-S{\'a}nchez}, F., {Prieto}, M.\,A., {Mezcua}, M., {Davies}, R.\,I.,
  {Malkan}, M.\,A., {Elitzur}, M.: 2013,  ApJL 763, L1

\bibitem{murgia1999}
{Murgia}, M., {Fanti}, C., {Fanti}, R., {Gregorini}, L., {Klein}, U., {Mack},
  K.\,H., {Vigotti}, M.: 1999,  A\&A 345, 769--777

\bibitem{odea1991}
{O'Dea}, C.\,P., {Baum}, S.\,A., {Stanghellini}, C.: 1991,  ApJ 380, 66--77

\bibitem{orienti2007}
{Orienti}, M., {Dallacasa}, D., {Stanghellini}, C.: 2007,  A\&A 461, 923--929

\bibitem{orienti2006}
{Orienti}, M., {Morganti}, R., {Dallacasa}, D.: 2006,  A\&A 457, 531--536

\bibitem{parma1999}
{Parma}, P., {Murgia}, M., {Morganti}, R., {Capetti}, A., {de Ruiter}, H.\,R.,
  {Fanti}, R.: 1999,  A\&A 344, 7--16

\bibitem{stanghellini}
{Stanghellini}, C., {O'Dea}, C.\,P., {Dallacasa}, D., {Cassaro}, P., {Baum},
  S.\,A., {Fanti}, R., {Fanti}, C.: 2005,  A\&A 443, 891--902

\bibitem{struveco}
{Struve}, C., {Conway}, J.\,E.: 2010,  A\&A 513, A10

\bibitem{struve2012}
{Struve}, C., {Conway}, J.\,E.: 2012,  A\&A 546, A22

\bibitem{tingay1997}
{Tingay}, S.\,J., {Jauncey}, D.\,L., {Reynolds}, J.\,E., {Tzioumis}, e.\,a.:
  1997,  AJ 113, 2025--2030

\bibitem{tingay2015}
{Tingay}, S.\,J., {Macquart}, J.\,P., {Collier}, J.\,D., {Rees}, G.,
  {Callingham}, J.\,R., {Stevens}, e.\,a.: 2015,  AJ 149, 74

\bibitem{vermeulen}
{Vermeulen}, R.\,C., {Pihlstr{\"o}m}, Y.\,M., {Tschager}, W., {de Vries},
  W.\,H., {Conway}, J.\,E., {Barthel}, e.\,a.: 2003,  A\&A 404, 861--870

\bibitem{veron}
{Veron-Cetty}, M.\,P., {Woltjer}, L., {Ekers}, R.\,D., {Staveley-Smith}, L.:
  1995,  A\&A 297, L79

\bibitem{willett}
{Willett}, K.\,W., {Stocke}, J.\,T., {Darling}, J., {Perlman}, E.\,S.: 2010,
  ApJ 713, 1393--1412

\end{thebibliography}
%

\end{document}